\documentstyle[prl,floats,aps,12pt]{revtex}

\begin{document}
\draft


\title{Harmonic-gauge dipole metric perturbations for weak-field 
circular orbits in Schwarzschild spacetime}
\author{Amos Ori}
\address{Department of Physics, Technion---Israel Institute of Technology,
Haifa, 32000, Israel}
\date{\today}
\maketitle

\begin{abstract}
We calculate the harmonic-gauge even $l=1$ mode of the linear metric
perturbation (MP) produced by a particle in a weak-field circular orbit
around a Schwarzschild black hole (BH). We focus on the Newtonian limit,
i.e. the limit in which the mass $M$ of the central BH approaches zero
(while fixing the orbital radius and the small-object mass), and obtain
explicit expressions for the MP in this limit. We find that the MP are
anomalous in this limit, namely, they do not approach their standard,
Coulomb-like, flat-space values. Instead, the MP grows on approaching the
BH, and this growth becomes worse as $M$ decreases. This anomalous behavior
leads to some pathologies which we briefly discuss. We also derive here the
next-order correction (in the orbital frequency $\Omega $) to the MP.

\end{abstract}


\vspace{2ex}
]


\section{Introduction}

One of the most useful gauges for analyzing linear metric perturbations (MP) 
$h_{\mu \nu }$ over a given background metric $g_{\mu \nu }$ is the \textit{%
harmonic gauge}. This gauge may be regarded as the gravitational analog of
the electromagnetic Lorentz gauge. The linearized Einstein equations in this
gauge reduce to a hyperbolic system [Eq. (\ref{einstein}) below], which
allows an elegant formulation of the initial-value problem. Also, the
phenomenon of gravitational self force \cite{mst}\cite{qw} was originally
formulated within this gauge. In this paper we shall show that the harmonic
gauge behaves in an unusual manner when a source is present in the
weak-field region of a Schwarzschild black hole (BH). \cite{dp}

The harmonic gauge is defined by the condition 
\begin{equation}
\bar{h}_{\,\,\,\,\,\,;\nu }^{\mu \nu }=0,  \label{cond}
\end{equation}
where $\bar{h}_{\mu \nu }$ is the trace-reversed MP, 
\[
\bar{h}_{\mu \nu }\equiv h_{\mu \nu }-\frac{1}{2}g_{\mu \nu }h_{\lambda
}^{\;\lambda }. 
\]
The linearized Einstein equations then reduce to 
\begin{equation}
\bar{h}_{\mu \nu ;\lambda }^{\;\;\;\;\;;\lambda }+2R_{\alpha \mu \beta \nu }%
\bar{h}^{\alpha \beta }=-16\pi T_{\mu \nu },  \label{einstein}
\end{equation}
where $R_{\alpha \mu \beta \nu }$ is the Riemann tensor associated with the
background metric $g_{\mu \nu }$.

We shall consider the gravitational perturbation produced by a point
particle of mass $\mu $. The simplest such problem is that of a static
particle in flat space. In this case the trace-reversed harmonic-gauge MP
takes the simple form 
\[
\bar{h}_{tt}=\frac{4\mu }{d}, 
\]
where $d$ is the distance from the particle (in the latter's rest frame),
with all other components of $\bar{h}_{\mu \nu }$ vanishing. This yields 
\begin{equation}
h_{\mu \nu }=\frac{2\mu }{d}\delta _{\mu \nu },  \label{flath}
\end{equation}
(in Cartesian coordinates), where $\delta _{\mu \nu }$ is $1$ for $\mu =\nu $
and zero otherwise.

Suppose now that the background geometry includes a BH of mass $M$, located
at a distance $R$ from the particle. Obviously the flat-space solution (\ref
{flath}) will no longer be valid. In this paper we shall address the
following question: What will happen to the harmonic-gauge functions $h_{\mu
\nu }(r^{\alpha })$ at the Newtonian limit where $M$ is taken to be
arbitrarily small while $R$ and the evaluation point $r^{\alpha }$ are kept
fixed? Will they approach the flat-space solution (\ref{flath})? We shall
analyze this problem (specifically for circular orbits in Schwarzschild
spacetime), and show they do not.

The same question may be asked for various kinds of perturbation fields in a
weak-field BH spacetime, and also to various modes of such fields. Assume,
for example, that the point particle is electrically charged. Will the
electromagnetic field produced by the particle approach its simple Coulombic
flat-space form, in the presence of a distant BH, in the Newtonian limit $%
M\to 0$? We shall say that a field ''behaves normally'' in the
Newtonian limit if it does approach its flat-space value as $M\to 0$%
. If, on the other hand, the field (or a particular mode of the field)
approaches a different limit as $M\to 0$, we shall refer to this
field (or mode) as \textit{anomalous}. In this paper we shall show that the
polar $l=1$ mode of the harmonic-gauge MP is anomalous; and so is the
overall harmonic-gauge MP (i.e. the sum over all modes). As a consequence,
when a small BH is located at a large distance $R$ from a finite-mass
source, the harmonic-gauge MP is pathological in some sense, and should be
used with great care. In particular, if $\mu R>>M^{2}$ (even though $\mu <<M$
) the harmonic-gauge MP grows large (i.e. $h_{\mu \nu }>>1$) in the
neighborhood of the BH, and the linear perturbation analysis actually breaks
down. (Obviously, it is possible to use the linear perturbation analysis for
the MP, even in this case---but not in the harmonic gauge!)

On physical grounds, one may expect that all locally-measurable quantities
will behave in a normal manner as $M\to 0$: It is unlikely that a
local measurement will be significantly affected by the presence of a
distant BH if the latter's mass is arbitrarily small. Therefore, the
anomalous behavior of the harmonic-gauge MP should be attributed to a bad
choice of gauge. That is, the harmonic gauge is advantageous in many cases,
but it becomes problematic (or even invalid, as in the case $\mu R>>M^{2}$)
in some weak-field situations.

The motivation for this work came from a recent analysis by Pfenning and
Poisson \cite{pf}. They calculated the gravitational self force (as well as
the electromagnetic and scalar self-forces) acting on a point particle in a
weak-field curved spacetime. Their analysis, which was based on the harmonic
gauge, indicated the crucial role of the $l=1$ mode in understanding the
(weak-field) gravitational self force, and hinted that this mode might
behave anomalously in the weak-field limit. Here we confirm that this is
indeed the case.

Let us now formulate our problem in a more precise form. We shall consider a
small particle of mass $\mu $ moving on a circular geodesic orbit around a
Schwarzschild BH of mass $M>>\mu $. We use the Schwarzschild coordinates for
the background, 
\[
ds^{2}=-(1-2M/r)dt^{2}+(1-2M/r)^{-1}dr^{2}+r^{2}(d\theta ^{2}+\sin
^{2}\theta d\varphi ^{2}). 
\]
The circular orbit is assumed to be equatorial, i.e. $\theta =\pi /2$, and
its radius is $r=R$. The two constants of motion are denoted $E\equiv -u_{t}$
and $L\equiv u_{\varphi }$. The orbital frequency is 
\[
\Omega \equiv \frac{d\varphi }{dt}=(M/R^{3})^{1/2}. 
\]
We shall consider a weak-field situation, i.e. $R>>M$, and analyze the even $%
l=1$ mode of the harmonic-gauge $h_{\mu \nu }$. Particularly we shall
calculate the limit $M\to 0$ (for fixed radius $R$ and location $%
r^{\alpha }$) of this mode. We shall show that this limit does not agree
with the corresponding flat-space expression; that is, the even $l=1$ mode
of the harmonic-gauge MP is anomalous.

We then go one step forward and calculate the first-order correction to this
mode in the small parameter $R\Omega =(M/R)^{1/2}$.

\subsection{The general method}

In our circular-orbit problem the energy-momentum source is restricted to a
delta-function distribution at $r=R$. At both sides $r<R$ and $r>R$ the MP
is source-less. We denote the harmonic-gauge vacuum MP at $r<R$ and $r>R$ by 
$h_{\mu \nu }^{(-)}$ and $h_{\mu \nu }^{(+)}$, respectively.

To construct the desired harmonic-gauge functions $h_{\mu \nu }^{(\pm )}$ we
proceed as follows. According to an analysis by Wald \cite{wald}, concerning
the $l=0,1$ modes of the gravitational perturbation over the Schwarzschild
geometry, if in some range $R_{1}<r<R_{2}$ of the perturbed Schwarzschild
background the gravitational perturbation is sourceless, then a gauge exists
in which the even $l=1$ part of the MP vanishes throughout this range. (In
fact Wald's analysis is more general, and it addresses the Kerr case as
well.) Thus, inside this range the even $l=1$ MP may be described in terms
of some even $l=1$ gauge vector $\xi _{\mu }$ through 
\begin{equation}
h_{\mu \nu }=-(\xi _{\mu ;\nu }+\xi _{\nu ;\mu }).  \label{trans}
\end{equation}
If we further impose the condition 
\begin{equation}
\Box \xi _{\mu }\equiv \xi _{\mu ;\lambda }^{\;\;\;;\lambda }=0,  \label{box}
\end{equation}
then the harmonic gauge condition (\ref{cond}) is satisfied by $h_{\mu \nu }$%
. The wave equation (\ref{einstein}) will then be satisfied too, because the
derived $h_{\mu \nu }$ is just a pure-gauge perturbation. Thus, the problem
of constructing the even $l=1$ vacuum harmonic-gauge MP reduces to that of
finding the even $l=1$ solutions to Eq. (\ref{box}). We shall refer to such
solutions of Eq. (\ref{box}) as ''homogeneous gauge vectors''.

Applying this procedure to each of the ranges $r<R$ and $r>R$, we conclude
that the two vacuum harmonic MP solutions $h_{\mu \nu }^{(-)}$ and $h_{\mu
\nu }^{(+)}$ may be derived through Eq. (\ref{trans}) from two even $l=1$
gauge vectors that solve Eq. (\ref{box}) (one in each range), which we
denote $\xi _{\mu }^{(-)}$ and $\xi _{\mu }^{(+)}$, respectively.

We emphasize that there does not exist a global gauge vector $\xi _{\mu
}^{(global)}$ in the entire domain $r>2M$ such that $h_{\mu \nu }$ is
globally derived from it via Eq. (\ref{trans}). In other words, the even $%
l=1 $ metric is trivial (i.e. may be annihilated by a gauge transformation)
locally in $r<R$ and separately in $r>R$, but not in the entire domain $r>2M$%
.

In principle the construction of the harmonic MP consists of two main
stages: (I) Constructing the general homogeneous gauge vectors, namely the
general even $l=1$ solutions to Eq. (\ref{box}); and (II) Finding $\xi _{\mu
}^{(-)}$ and $\xi _{\mu }^{(+)}$, i.e. the \textit{specific}, actual gauge
vectors at $r<R$ and $r>R$, respectively. The choice of the actual gauge
vectors $\xi _{\mu }^{(\pm )}$ should be made, in principle, by first
constructing the general vacuum solutions $h_{\mu \nu }^{(\pm )}$ from the
general homogeneous gauge-vector solution at both sides, and then imposing
the following conditions on $h_{\mu \nu }^{(\pm )}$: (a) Appropriate
asymptotic behavior at the event horizon (EH) and at infinity, (b)
continuity at $r=R$, and (c) A proper discontinuity of the $r$ -derivatives
of $h_{\mu \nu }^{(\pm )}$ at $r=R$, as determined from the energy-momentum
distribution there through the field equation (\ref{einstein}).

In practice, the condition (a) is most easily implemented at the level of $%
\xi .$ As can be easily verified, if a gauge field $\xi _{\mu }$ is regular
at the EH, the MP constructed from it via Eq. (\ref{trans}) is regular at
the EH too. The situation at the limit $r\to \infty $ is similar: If
a gauge field $\xi _{\mu }$ vanishes at infinity, the corresponding MP will
vanish at this limit too.

To the best of our knowledge the exact explicit solutions to Eq. (\ref{box})
(for $M\neq 0$ and frequencies $\omega \neq 0$) are not known. Here,
however, we are only interested in the small-$M$ limit, i.e. the situation
in which $M<<R$. We shall thus calculate the MP at the leading order $
(M/R)^{0},(R\Omega )^{0}$, and also at the next order $(R\Omega )^{1}$. To
this end we shall only need an approximate expression for the gauge vector,
accurate to a certain order in the small parameters $M/R$ and $R\Omega $.
Such an approximate expression can easily be constructed.

In section II we shall construct the general homogeneous even $l=1$ gauge
vector, at the leading order [i.e. order $(M/R)^{0},(R\Omega )^{0}$]. Then
in section III we correct this homogeneous solution at orders $(M/R)^{1}$
and $(R\Omega )^{2}$ (these turn out to be the orders required for our
analysis). In section IV we consider the energy-momentum source associated
with the massive particle and construct its even $l=1$ mode. Then in section
V we use this source, combined with the general homogeneous solutions
derived in section III, to construct the actual gauge vector $\xi _{\mu
}^{(\pm )}$ at the two vacuum regions $r<R$ and $r>R$. From this gauge
vector we then derive in section VI the harmonic-gauge even $l=1$ MP
functions $h_{\mu \nu }^{(\pm )}$ ---first at the leading order $%
(M/R)^{0}=(R\Omega )^{0}$, and then at the next order $(R\Omega )^{1}$. Then
in section VI we explore the overall harmonic-gauge MP, i.e. the sum over
the contributions of all modes, at the limit $M\to 0$. We thus
express the overall MP as the sum of the flat-space term $h_{\mu \nu }^{0}$
and an anomalous term $\Delta h_{\mu \nu }$ (independent of $M$). Finally,
in section VIII we briefly discuss the results. In particular we show that
the harmonic-gauge MP becomes $>>1$ (and hence invalid as a linear
perturbation) if the particle is located too far from the BH. In Appendix A
we present five of the six exact even, $l=1$, static homogeneous solutions
to Eq. (\ref{box}) in the Schwarzschild background (the sixth solution,
which we denote $\xi _{\mu }^{(C+)}$, is not presented as it is too long).

\section{Homogeneous gauge-vector solutions: Leading order}

\subsection{Flat-space static solutions}

We start by exploring the static even $l=1$ solutions of Eq. (\ref{box}) in
flat space. For each azimuthal number $m$ ($m$ gets the three values $1\leq
m\leq 1$) there are six such solutions, which are naturally divided into
three pairs. We denote them $\xi _{\mu }^{(A\pm )},\xi _{\mu }^{(B\pm )},\xi
_{\mu }^{(C\pm )},$ where the ''+'' label refers to solutions that vanish at
infinity, and the ''-'' label refers to solutions that are regular at the
origin.

We first present the six $m=0$ solutions, which are axially-symmetric. The
first pair is the ''scalar mode'', 
\[
\xi _{\mu }^{(A\pm )}=\phi _{;\mu }^{(\pm )}\;, 
\]
where 
\[
\phi ^{(-)}=r\cos \theta \;,\;\phi ^{(+)}=r^{-2}\cos \theta . 
\]
[Note that $\phi ^{(\pm )}$ are just the two axially-symmetric $l=1,m=0$
solutions to the scalar wave equation.] The next pair is the ''electric
mode'' 
\begin{equation}
\xi _{t}^{(B-)}=r\cos \theta \;,\;\xi _{t}^{(B+)}=r^{-2}\cos \theta ,
\end{equation}
\[
\xi _{r}^{(B\pm )}=\xi _{\theta }^{(B\pm )}=\xi _{\varphi }^{(B\pm )}=0. 
\]
The last pair is 
\[
\xi _{r}^{(C-)}=r^{2}\cos \theta \;,\;\xi _{r}^{(C+)}=r^{-1}\cos \theta , 
\]
\[
\xi _{\theta }^{(C-)}=\frac{1}{2}r^{3}\sin \theta \;,\;\xi _{\theta
}^{(C+)}=-\sin \theta , 
\]
\[
\xi _{\varphi }^{(C\pm )}=\xi _{t}^{(C\pm )}=0. 
\]

The related $m=\pm 1$ modes are obtained by a straightforward rotation of
the $m=0$ solution. For our purpose it is convenient to consider the real
modes (i.e. $\sin $ and $\cos $ functions of $\varphi $ rather than complex
exponents). In one such combination of the $m=\pm 1$ azimuthal numbers (the
combination which is mostly relevant to the analysis below) the six
solutions are as follows. The ''scalar mode'' is 
\begin{equation}
\xi _{\mu }^{(A\pm )}=\phi _{;\mu }^{(\pm )}\;  \label{statica1}
\end{equation}
with 
\begin{equation}
\phi ^{(-)}=r\sin \theta \cos \varphi \;,\;\phi ^{(+)}=r^{-2}\sin \theta
\cos \varphi ,  \label{statica2}
\end{equation}
the ''electric mode'' is \footnote{%
Here and throughout this paper we take this specific phase ($\sin \varphi $
as opposed to the phase factor $\cos \varphi $ used above for the scalar
mode) for the ''electric mode'', because it is this phase which emerges from
the matching to the source at $r=R$, (see section V).} 
\begin{equation}
\xi _{t}^{(B-)}=r\sin \theta \sin \varphi \;,\;\xi _{t}^{(B+)}=r^{-2}\sin
\theta \sin \varphi  \label{staticb1}
\end{equation}
\begin{equation}
\xi _{r}^{(B\pm )}=\xi _{\theta }^{(B\pm )}=\xi _{\varphi }^{(B\pm )}=0,
\label{staticb2}
\end{equation}
and the third pair is 
\begin{equation}
\xi _{r}^{(C-)}=r^{2}\sin \theta \cos \varphi \;,\;\xi
_{r}^{(C+)}=r^{-1}\sin \theta \cos \varphi ,  \label{staticc1}
\end{equation}
\begin{equation}
\xi _{\theta }^{(C-)}=-\frac{1}{2}r^{3}\cos \theta \cos \varphi \;,\;\xi
_{\theta }^{(C+)}=\cos \theta \cos \varphi ,  \label{staticc2}
\end{equation}
\begin{equation}
\xi _{\varphi }^{(C-)}=\frac{1}{2}r^{3}\sin \theta \sin \varphi \;,\;\xi
_{\varphi }^{(C+)}=-\sin \theta \sin \varphi ,  \label{staticc3}
\end{equation}
\begin{equation}
\xi _{t}^{(C\pm )}=0.  \label{staticc4}
\end{equation}
The other set of six $m=\pm 1$ solutions is now obtained by replacing $%
\varphi \to \varphi +\pi /2$.

\subsection{Slowly-rotating weak-field solutions}

We now return to the problem of weak-field circular orbits in Schwarzschild.
Our goal at this stage is to construct the relevant homogeneous solutions,
at the leading order in $M$ and $\Omega $, namely at orders $(M/R)^{0}$ and $%
(R\Omega )^{0}$. At this order, all radial functions simply take their above
static flat-space form. Since the orbit is equatorial, the $m=0$ solutions
turn out to be irrelevant (because the even $l=1$ part of the source term
does not include $m=0$ modes---see section IV). The $m=\pm 1$ solutions for
the slowly-rotating modes are obtained from Eqs. (\ref{statica1}-\ref
{staticc4}) by simply replacing $\varphi \to \varphi -\Omega t$. The
set of six leading-order solutions is again divided into three pairs. The
''scalar mode'' is 
\begin{equation}
\xi _{\mu }^{(A\pm )}=\phi _{;\mu }^{(\pm )}\;,  \label{leada1}
\end{equation}
with 
\begin{equation}
\phi ^{(-)}=r\sin \theta \cos (\varphi -\Omega t)\;,\;\phi ^{(+)}=r^{-2}\sin
\theta \cos (\varphi -\Omega t).  \label{leada2}
\end{equation}
The ''electric mode'' is 
\begin{equation}
\xi _{t}^{(B-)}=r\sin \theta \sin (\varphi -\Omega t)\;,\;\xi
_{t}^{(B+)}=r^{-2}\sin \theta \sin (\varphi -\Omega t)  \label{leadb1}
\end{equation}
with 
\begin{equation}
\xi _{r}^{(B\pm )}=\xi _{\theta }^{(B\pm )}=\xi _{\varphi }^{(B\pm )}=0.
\label{leadb2}
\end{equation}
The last pair is 
\begin{equation}
\xi _{r}^{(C-)}=r^{2}\sin \theta \cos (\varphi -\Omega t)\;,\;\xi
_{r}^{(C+)}=r^{-1}\sin \theta \cos (\varphi -\Omega t),  \label{leadc1}
\end{equation}
\begin{equation}
\xi _{\theta }^{(C-)}=-\frac{1}{2}r^{3}\cos \theta \cos (\varphi -\Omega
t)\;,\;\xi _{\theta }^{(C+)}=\cos \theta \cos (\varphi -\Omega t),
\label{leadc2}
\end{equation}
\begin{equation}
\xi _{\varphi }^{(C-)}=\frac{1}{2}r^{3}\sin \theta \sin (\varphi -\Omega
t)\;,\;\xi _{\varphi }^{(C+)}=-\sin \theta \sin (\varphi -\Omega t),
\label{leadc3}
\end{equation}
\begin{equation}
\xi _{t}^{(C\pm )}=0.  \label{leadc4}
\end{equation}
As before, the other six $m=\pm 1$ solutions are obtained by transforming $%
\varphi \to \varphi +\pi /2$. (However, this second set of $m=\pm 1$
solutions will not be involved in the construction below, for the same
reason as for the set of $m=0$ solutions.)

Note that we do not expand the functions $\cos (\varphi -\Omega t)$ or $\sin
(\varphi -\Omega t)$ in $\Omega $, for the following obvious reason:
Although $\Omega $ is assumed to be small (compared to e.g. $1/R$ and $1/r$
), $\Omega t$ may be arbitrarily large. Only the radial functions are
expanded in $\Omega $ ---and in this section we consider the leading order,
which is independent of $\Omega $ (or $M$).

Our starting point in constructing the above leading-order solutions was the
set of flat-space static solutions. Alternatively, one may start from the
exact static solutions on the Schwarzschild background, given in Appendix A,
and then take the limit $M\to 0$. This leads to the same results,
Eqs. (\ref{leada1}-\ref{leadc4}). In the static Schwarzschild case, too,
there are three ''-'' solutions that are regular at the EH, and three ''+''
solutions that vanish at infinity. It is important to note that at the limit 
$M\to 0$ the Schwarzschild's ''+'' and ''-'' solutions properly
approach the corresponding flat-space ''+'' and ''-'' solutions. That is,
the Schwarzschild's $l=1$ gauge fields are non-anomalous at the limit $%
M\to 0$ (this is to be contrasted with the corresponding MP
functions, which are anomalous, as we show below).

\section{Homogeneous gauge-vector solutions: Higher-order corrections}

In the construction of the actual gauge-vector solution, some of the above
homogeneous solutions must be calculated to higher order in the small
parameters $M/r$ and $r\Omega $. The basic reason is the following: As we
show below [see e.g. Eq. (\ref{coefin})], in constructing the actual
gauge-vector solution $\xi _{\mu }^{(\pm )}$, the homogeneous solution $\xi
_{\mu }^{(A-)}$ enters with a prefactor proportional to $(M/R)^{-1}$; and
similarly, $\xi _{\mu }^{(B-)}$ enters with a prefactor proportional to $%
(R\Omega )^{-1}$. [All other homogeneous solutions enter with a prefactor
proportional to $(R\Omega )^{0}$ or even smaller]. Note the following
anomalous situation: Whereas the harmonic MP scales at the leading order as $%
(M/R)^{0}$, the actual gauge vector $\xi _{\mu }$ scales as $(M/R)^{-1}$. In
order to obtain the MP at the leading-order $(M/R)^{0}$, we must determine
the actual gauge vector $\xi _{\mu }$ at the same order. This requires us to
determine $\xi _{\mu }^{(A-)}$ with accuracy $(M/r)^{1}$ and $(r\Omega )^{2}$%
.

Furthermore, we would like to go one step forward and to evaluate the
correction to the MP proportional to $(R\Omega )^{1}$. This requires one to
determine $\xi _{\mu }$ with the same accuracy $(R\Omega )^{1}$. In turn,
this requires us to determine $\xi _{\mu }^{(A-)}$ at order $(r\Omega )^{3}$
[this also includes corrections of the type $(r\Omega )(M/r)=(M\Omega )$], $%
\xi _{\mu }^{(B-)}$ at order $(r\Omega )^{2}$ and $(M/r)$, and all other
homogeneous solutions at order $(r\Omega )^{1}$.

In practice, the corrections to the radial functions of the homogeneous
gauge vector only emerge in even powers of $r\Omega $ (but any power of $M/r$
), and no odd powers of $r\Omega $ are involved (this law brakes at some
order in $R\Omega $ which is beyond our present goal). This implies the
following:

To calculate the MP at order $(R\Omega )^{0}$, we need to evaluate $\phi
^{(-)}$ (the generating function for $\xi _{\mu }^{(A-)}$) at order $%
(r\Omega )^{2}$ and $(M/r)^{1}$, and all other homogeneous solutions at the
leading order $(r\Omega )^{0}$ and $(M/r)^{0}$. And in order to get one
extra order (i.e. order $R\Omega $) in the MP, all we need is to calculate $%
\xi _{\mu }^{(B-)}$ , too, at order $(r\Omega )^{2}$ and $(M/r)^{1}$.

Summarizing these considerations, since our ultimate goal is to calculate
the MP at accuracy $(R\Omega )^{1}$, we need to calculate $\phi ^{(-)}$ and $%
\xi _{\mu }^{(B-)}$ at order $(r\Omega )^{2}$ and $(M/r)^{1}$, and all other
inhomogeneous solutions at the leading order $(r\Omega )^{0}$ and $(M/r)^{0}$
.

Note that up to the order required here [i.e. $(r\Omega )^{2}$ and $%
(M/r)^{1} $] there is no coupling between the effects of rotation (i.e. $%
\Omega \neq 0$) and curvature (i.e. $M\neq 0$). These effects first couple
at order $(r\Omega )(M/r)=M\Omega =(r\Omega )^{3}$, which is not required in
our analysis. This situation allows the following simple procedure. The $%
O(r\Omega )^{2}$ corrections are obtained by exploring the flat-space $%
\Omega \neq 0$ radial functions; And the $O(M/r)$ corrections are determined
from the exact static homogeneous radial functions in the $M\neq 0$
Schwarzschild case. Note that both kinds of corrections are only required
for the two homogeneous solutions $\xi _{\mu }^{(A-)}$ (or, in practice, $%
\phi ^{(-)}$) and $\xi _{\mu }^{(B-)}$. In what follows we shall describe in
more detail these two types of corrections.

\subsection{Rotation-induced corrections}

Consider first the scalar solution $\phi ^{(-)}$. We need to explore here
the homogeneous solutions to the scalar $l=1$ radial equation, for a mode $%
\omega =\Omega \neq 0$ [specifically we consider here the solution with the
angular dependence given in Eq. (\ref{leada2}) above]. We only need here the
''-'' solution, i.e. the one regular at the origin. This solution takes the
simple exact form 
\[
\phi ^{(-)}=3\left[ \frac{\sin (r\Omega )-r\Omega \cos (r\Omega )}{(r\Omega
)^{3}}\right] r\sin \theta \cos (\varphi -\Omega t) 
\]
(the factor $3$ was introduced here in order for $\phi ^{(-)}$ to comply
with the above leading-order expression for $\phi ^{(-)}$). In our analysis
we shall only need the corrections to the radial function up to order $%
(r\Omega )^{2}$, which read 
\[
\phi ^{(-)}\cong \left[ 1-\frac{(r\Omega )^{2}}{10}\right] r\sin \theta \cos
(\varphi -\Omega t). 
\]

Next consider the exact solution $\xi _{\mu }^{(B-)}$ for a flat space but
with $\Omega \neq 0$. The equation imposed on $\xi _{\mu }^{(B-)}$ is 
\[
\Box \xi _{\mu }^{(B-)}=0. 
\]
In flat space, each Cartesian component satisfies this box equation
separately. Therefore, an exact solution to this equation is 
\begin{equation}
\xi _{\mu }^{(B-)}=3\delta _{\mu }^{t}\left[ \frac{\sin (r\Omega )-r\Omega
\sin (r\Omega )}{(r\Omega )^{3}}\right] r\sin \theta \sin (\varphi -\Omega
t)\,.  \label{B1}
\end{equation}
We adopt this solution as the exact version of the leading-order expressions
(\ref{leadb1}-\ref{leadb2}) for $\xi _{\mu }^{(B-)}$.\footnote{%
In the static flat-space case $\xi _{\mu }^{(B\pm )}$ coincides with the $%
l=1 $ mode of the Lorentz-gauge electromagnetic four-potential (hence the
phrase ''electric mode''). In the $\Omega \neq 0$ case, $\xi _{\mu }^{(B-)}$
as defined here no longer agrees with the Lorentz-gauge electromagnetic
four-potential, as it is not divergence-free. Nevertheless, Eq. (\ref{B1})
is a legitimate extension of Eqs. (\ref{staticb1}-\ref{staticb2}) to the $%
\Omega \neq 0$ case, which considerably simplifies the calculations.} The
decomposition of the radial function in $(r\Omega )$ then yields 
\[
\xi _{\mu }^{(B-)}\cong \delta _{\mu }^{t}\left[ 1-\frac{(r\Omega )^{2}}{10}%
\right] r\sin \theta \sin (\varphi -\Omega t). 
\]

\subsection{Curvature-induced corrections}

To obtain the $M/r$ corrections we explore the exact solutions for the
homogeneous, static, $l=1$ gauge vector in Schwarzschild spacetime (see
Appendix A). For the scalar mode $\phi ^{(-)}$ we have the simple exact
solution 
\[
\phi ^{(-)}=(r-M)\sin \theta \cos (\varphi -\varphi _{0})\,. 
\]
For $\xi _{\mu }^{(B-)}$ we have the exact solution 
\[
\xi _{\mu }^{(B-)}=\delta _{\mu }^{t}(r-2M)\sin \theta \sin (\varphi
-\varphi _{0})\,. 
\]

\subsection{The total corrections}

Summing the above corrections to the radial functions of $\phi ^{(-)}$ and $%
\xi _{\mu }^{(B-)}$, which emerge from both the rotation and the
nonvanishing of $M$ (and expanding them in the small parameters $r\Omega $
and $M/r$), we obtain 
\begin{equation}
\phi ^{(-)}\cong \left[ 1-\frac{M}{r}-\frac{(r\Omega )^{2}}{10}\right] r\sin
\theta \cos (\varphi -\Omega t)  \label{Aexact}
\end{equation}
and 
\begin{equation}
\xi _{\mu }^{(B-)}\cong \delta _{\mu }^{t}\left[ 1-\frac{2M}{r}-\frac{%
(r\Omega )^{2}}{10}\right] r\sin \theta \sin (\varphi -\Omega t).
\label{Bexact}
\end{equation}
These corrected expressions for $\phi ^{(-)}$ and $\xi _{\mu }^{(B-)}$ will
be used in section V, together with the leading-order expressions (\ref
{leada1}-\ref{leadc4}) for $\phi ^{(+)},\xi _{\mu }^{(B+)}$ and $\xi
_{\varphi }^{(C\pm )}$, in the construction of the actual gauge vector $\xi
_{\mu }^{(\pm )}$.

\section{The energy-momentum source}

To construct the actual gauge vector at both sides of $r=R$ we need the $l=1$
part of the energy-momentum distribution at $r=R$.

The particle's equatorial circular orbit is given by 
\[
r=R,\;\theta =\pi /2,\;\varphi =u^{\varphi }\tau ,\;t=u^{t}\tau . 
\]
For this kind of orbit $u^{\varphi }$ and $u^{t}$ are constants that depend
on $M$ and $R$, and satisfy $\Omega =u^{\varphi }/u^{t}$ (without loss of
generality we have set $t=\varphi =0$ at $\tau =0$). The energy-momentum
tensor is 
\begin{eqnarray}
T_{\mu \nu } &=&\mu \int (-g)^{-1/2}u_{\mu }u_{\nu }\delta (r-R)\delta
(\theta -\pi /2)\delta (\varphi -u^{\varphi }\tau )\delta (t-u^{t}\tau )d\tau
\nonumber \\
&=&\mu (R^{2}\sin \theta u^{t})^{-1}\delta (r-R)\delta (\theta -\pi
/2)\delta (\varphi -\Omega t)u_{\mu }u_{\nu }.  \label{T}
\end{eqnarray}
This tensor is now expanded in tensorial harmonics $Y_{\mu \nu }^{(i)lm}$.
Throughout this paper we use the notation introduced in Ref. \cite{barack}
for this decomposition. For a general tensor $T_{\mu \nu }(x^{\alpha })$ the
tensorial mode decomposition takes the form 
\[
T_{\mu \nu }(r,t,\theta ,\varphi )=\sum_{l=0}^{\infty
}\sum_{m=-l}^{l}\sum_{i=1}^{10}Y_{\mu \nu }^{(i)lm}(\theta ,\varphi
)T^{(i)lm}(r,t). 
\]
The ''coefficients'' $T^{(i)lm}$ are given by 
\begin{equation}
T^{(i)lm}(r,t)=k^{(i)}\int \eta ^{\mu \alpha }\eta ^{\nu \beta }[Y_{\alpha
\beta }^{(i)lm}(\theta ,\varphi )]^{*}T_{\mu \nu }(r,t,\theta ,\varphi )\sin
\theta d\theta d\varphi ,  \label{Tcoef}
\end{equation}
where 
\[
\eta ^{\mu \alpha }\equiv diag[-1,1,r^{-2},(r\sin \theta )^{-2}], 
\]
and $k^{(i)}$ is either $+1$ or $-1$. For a general $l$ the harmonics $%
i=1,...,7$ are even and $i=8,...,10$ are odd, but for $l=1$ the harmonic $%
i=7 $ vanishes identically. Since $u_{r}=0$, the source term (\ref{T}) does
not have any $T_{r\mu }$ component, and consequently the harmonics $i=2,3,5$
vanish (see Eq. (13) in Ref. \cite{barack}). We are thus left with three
even harmonics, $i=1,4,6$: 
\begin{eqnarray*}
T_{\mu \nu }^{l=1,even} &=&\sum_{m=-1}^{1}\left( Y_{\mu \nu
}^{(1)1m}T^{(1)1m}+Y_{\mu \nu }^{(4)1m}T^{(4)1m}+Y_{\mu \nu
}^{(6)1m}T^{(6)1m}\right) \\
&\equiv &Z_{\mu \nu }^{(1)}+Z_{\mu \nu }^{(4)}+Z_{\mu \nu }^{(6)}.
\end{eqnarray*}
(Note that $k=1$ for $i=1,6$ and $k=-1$ for $i=4$.) We shall now proceed to
calculate the contributions of these three harmonics.

The harmonic $i=1$ is 
\[
Y_{\mu \nu }^{(1)lm}=Y^{lm}\delta _{\mu }^{t}\delta _{\nu }^{t}, 
\]
and a straightforward calculation yields 
\[
T^{(1)lm}=\mu E^{2}(R^{2}u^{t})^{-1}\delta (r-R)[Y^{lm}(\theta =\pi
/2,\varphi =\Omega t)]^{*}. 
\]
Since $Y^{1,0}\propto \cos \theta $ vanishes at $\theta =\pi /2$, the only
relevant spherical harmonics are $m=\pm 1$, given by \footnote{%
We adopt here the notation used in Ref. \cite{ll} for the scalar spherical
harmonics.} 
\[
Y^{1,\pm 1}(\theta ,\varphi )=\mp i\sqrt{3/8\pi }\sin \theta e^{\pm i\varphi
}. 
\]
One finds 
\[
Y_{\mu \nu }^{(1)1,\pm 1}T^{(1)1,\pm 1}=(3/8\pi )\mu
E^{2}(R^{2}u^{t})^{-1}\delta (r-R)\sin \theta e^{\pm i(\varphi -\Omega
t)}\delta _{\mu }^{t}\delta _{\nu }^{t}\,, 
\]
hence 
\[
Z_{\mu \nu }^{(1)}=(3/4\pi )\mu E^{2}(R^{2}u^{t})^{-1}\delta (r-R)\sin
\theta \cos (\varphi -\Omega t)\delta _{\mu }^{t}\delta _{\nu }^{t}\,. 
\]

The $i=4$ tensor harmonic is 
\[
Y_{\mu \nu }^{(4)lm}=\frac{ir}{\sqrt{2l(l+1)}}[Y_{,\theta }^{lm}(\delta
_{\mu }^{t}\delta _{\nu }^{\theta }+\delta _{\mu }^{\theta }\delta _{\nu
}^{t})+Y_{,\varphi }^{lm}(\delta _{\mu }^{t}\delta _{\nu }^{\varphi }+\delta
_{\mu }^{\varphi }\delta _{\nu }^{t})]. 
\]
In calculating $T^{(i)lm}$ through Eq. (\ref{Tcoef}), the term $\propto
Y_{,\theta }^{lm}$ of $Y_{\mu \nu }^{(4)lm}$ multiplies $T_{t\theta }$,
which identically vanishes for an equatorial orbit. The only contribution to 
$T^{(i)lm}$ thus comes from the term proportional to $Y_{,\varphi }^{lm}$.
The latter vanishes for $m=0$, hence again there will only be contributions
from $m=\pm 1$. The explicit form of $Y_{\mu \nu }^{(4)lm}$ for these modes
is 
\begin{eqnarray*}
Y_{\mu \nu }^{(4)1,\pm 1} &=&\frac{r}{2}\sqrt{3/8\pi }e^{\pm i\varphi } \\
&&\times [\pm \cos \theta (\delta _{\mu }^{t}\delta _{\nu }^{\theta }+\delta
_{\mu }^{\theta }\delta _{\nu }^{t})+i\sin \theta (\delta _{\mu }^{t}\delta
_{\nu }^{\varphi }+\delta _{\mu }^{\varphi }\delta _{\nu }^{t})].
\end{eqnarray*}
After a straightforward calculation one finds 
\[
T^{(4)1,\pm 1}=i\sqrt{3/8\pi }\mu EL(R^{3}u^{t})^{-1}\delta (r-R)e^{\mp
i\Omega t}, 
\]
which yields 
\begin{eqnarray*}
Z_{\mu \nu }^{(4)} &=&\frac{-3\mu EL}{8\pi R^{2}u^{t}}\delta (r-R)\,\times \\
&&\left[ \cos \theta \sin (\varphi -\Omega t)(\delta _{\mu }^{t}\delta _{\nu
}^{\theta }+\delta _{\mu }^{\theta }\delta _{\nu }^{t})+\sin \theta \cos
(\varphi -\Omega t)(\delta _{\mu }^{t}\delta _{\nu }^{\varphi }+\delta _{\mu
}^{\varphi }\delta _{\nu }^{t})\right] .
\end{eqnarray*}

Finally, the $i=6$ harmonic is 
\[
Y_{\mu \nu }^{(6)lm}=\frac{r^{2}}{\sqrt{2}}Y^{lm}(\delta _{\mu }^{\theta
}\delta _{\nu }^{\theta }+\sin ^{2}\theta \delta _{\mu }^{\varphi }\delta
_{\nu }^{\varphi }). 
\]
Again, $T^{(6)1,\pm 1}$ will not get a contribution from the term $\propto
\delta _{\mu }^{\theta }\delta _{\nu }^{\theta }$ in $Y_{\mu \nu }^{(6)lm}$,
because $T_{\theta \theta }$ vanishes. The contribution from the other term $%
\propto \delta _{\mu }^{\varphi }\delta _{\nu }^{\varphi }$ will be
proportional to $T_{\varphi \varphi }$, which in turn is proportional to $%
u_{\varphi }^{2}\propto \Omega ^{2}$. Since our analysis is limited to the
order $\Omega ^{1}$, it will not include contributions from $i=6$.

In summary, at the desired accuracy ($\Omega ^{1}$), the energy-momentum
distribution is 
\begin{equation}
T_{\mu \nu }^{l=1,even}=Z_{\mu \nu }^{(1)}+Z_{\mu \nu }^{(4)}.
\label{Tfinal}
\end{equation}
Since we ignore here $O(M/R)$ correction terms, we may substitute $%
u^{t}\cong E\cong 1$ and $L\cong R^{2}\Omega $, yielding 
\begin{equation}
Z_{\mu \nu }^{(1)}=(3/4\pi )\mu R^{-2}\delta (r-R)\sin \theta \cos (\varphi
-\Omega t)\delta _{\mu }^{t}\delta _{\nu }^{t}  \label{Z1}
\end{equation}
and 
\begin{eqnarray*}
Z_{\mu \nu }^{(4)} &=&-(3/8\pi )\mu \Omega \delta (r-R)\,\times \\
&&\left[ \cos \theta \sin (\varphi -\Omega t)(\delta _{\mu }^{t}\delta _{\nu
}^{\theta }+\delta _{\mu }^{\theta }\delta _{\nu }^{t})+\sin \theta \cos
(\varphi -\Omega t)(\delta _{\mu }^{t}\delta _{\nu }^{\varphi }+\delta _{\mu
}^{\varphi }\delta _{\nu }^{t})\right] .
\end{eqnarray*}
Note that $Z_{\mu \nu }^{(1)}$ is of the leading order $\Omega ^{0}$ and $%
Z_{\mu \nu }^{(4)}$ is $O(\Omega ^{1}).$

\section{The actual gauge vector}

In both ranges $r<R$ and $r>R$ the perturbation is sourceless, hence in each
of these ranges the even $l=1$ mode may be described by a certain
homogeneous gauge vector [i.e. a certain even $l=1$ solution to the equation
(\ref{box})]. We denote this actual gauge vector by $\xi _{\mu }^{(-)}$ at $%
r<R$ and $\xi _{\mu }^{(+)}$ at $r<R$. Each of these vector fields may be
expressed as a certain combination of the basis homogeneous solutions
constructed in sections II and III. In order for $\xi _{\mu }^{(+)}$ to
vanish at $r\to \infty $, it must be composed of the three ''+''
solutions solely. Similarly, in order for $\xi _{\mu }^{(-)}$ to be regular
at the EH, it must be composed of the three ''-'' solutions solely. 
\footnote{%
Recall that the three approximate ''-'' solutions constructed above may be
obtained from the corresponding three static ''-'' solutions in
Schwarzschild [by taking the leading order(s) in $M/r$, and adding the
rotation-induced correction terms when required]; and the three ''-'' static
Schwarzschild's solutions are regular at the EH by construction (see
Appendix A).} Thus, $\xi _{\mu }^{(\pm )}$ must take the form 
\begin{equation}
\xi _{\mu }^{(\pm )}=a^{(\pm )}\xi _{\mu }^{(A\pm )}+b^{(\pm )}\xi _{\mu
}^{(B\pm )}+c^{(\pm )}\xi _{\mu }^{(C\pm )}.  \label{vecgen}
\end{equation}
The MP functions $h_{\mu \nu }^{(\pm )}$ at the two sides of $r=R$ are
constructed from $\xi _{\mu }^{(\pm )}$ via 
\begin{equation}
h_{\mu \nu }^{(\pm )}=-(\xi _{\mu ,\nu }^{(\pm )}+\xi _{\nu ,\mu }^{(\pm )}).
\label{hpm}
\end{equation}
The six coefficients $a^{(\pm )},b^{(\pm )},c^{(\pm )}$ are to be obtained
by imposing two matching conditions on the MP:

(i) $h_{\mu \nu }$ must be continuous at $r=R$: 
\[
h_{\mu \nu }^{(+)}=h_{\mu \nu }^{(-)}, 
\]

(ii) The radial derivatives $h_{\mu \nu ,r}$ will undergo a certain jump at $%
r=R$. This jump is conveniently described in terms of the trace-reversed MP, 
$\bar{h}_{\mu \nu }$, through the quantities 
\[
<\bar{h}_{\mu \nu ,r}>\equiv [\bar{h}_{\mu \nu ,r}^{(+)}-\bar{h}_{\mu \nu
,r}^{(-)}]_{r=R}\,. 
\]
The jump is to be determined from the even $l=1$ energy-momentum
distribution at $r=R$, Eq. (\ref{Tfinal}). Applying the linearized Einstein
equation (\ref{einstein}) to this thin shell (approximating $g^{rr}\cong 1$%
), one finds the required jump conditions: 
\begin{equation}
<\bar{h}_{tt,r}>=-12\mu R^{-2}\sin \theta \cos (\varphi -\Omega t)
\label{j1}
\end{equation}
at order $\Omega ^{0}$, and 
\begin{equation}
<\bar{h}_{t\theta ,r}>=<\bar{h}_{\theta t,r}>=6\mu \Omega \cos \theta \sin
(\varphi -\Omega t),  \label{j2}
\end{equation}
\begin{equation}
<\bar{h}_{t\varphi ,r}>=<\bar{h}_{\varphi t,r}>=6\mu \Omega \sin \theta \cos
(\varphi -\Omega t)  \label{j3}
\end{equation}
at order $\Omega ^{1}$, with all other components of $<\bar{h}_{\mu \nu ,r}>$
vanishing.

The calculation of the six coefficients $a^{(\pm )},b^{(\pm )},c^{(\pm )}$
thus proceeds as follows. One first applies Eq. (\ref{hpm}) to the actual
gauge vector written in the form (\ref{vecgen}), then apply the continuity
condition (i), and then the jump condition (ii). The calculation is lengthy
but straightforward, and we shall skip here the details. The values of the
above six coefficients are found to be

\begin{eqnarray}
a^{(-)} &=&\left( \frac{R}{M}-\frac{5}{2}\right) \,\mu ,\;b^{(-)}=\left( -2%
\frac{R}{M}+1\right) \Omega \mu \,,\;  \nonumber \\
c^{(-)} &=&\frac{4}{5R^{2}}\mu \,,\;  \label{coefin}
\end{eqnarray}
and 
\begin{equation}
a^{(+)}=0,\;b^{(+)}=-\frac{2R^{3}}{5}\Omega \mu \,,\;c^{(+)}=-2R\mu \,.
\label{coefout}
\end{equation}
The actual gauge vector at $r<R$ then becomes 
\[
\xi _{t}^{(-)}=\mu \left( -\frac{R}{M}r+\frac{r^{3}-15R^{2}r+30R^{3}}{10R^{2}%
}\right) \Omega \sin \theta \sin (\varphi -\Omega t)\;, 
\]
\[
\xi _{r}^{(-)}=\mu \left( \frac{R}{M}+\frac{r^{2}-5R^{2}}{2R^{2}}\right)
\sin \theta \cos (\varphi -\Omega t)\;, 
\]
\[
\xi _{\theta }^{(-)}=\mu \left( \frac{R}{M}r-\frac{r^{3}+5R^{2}r+2R^{3}}{%
2R^{2}}\right) \cos \theta \cos (\varphi -\Omega t)\;, 
\]
\[
\xi _{\varphi }^{(-)}=\mu \left( -\frac{R}{M}r+\frac{r^{3}+5R^{2}r+2R^{3}}{
2R^{2}}\right) \sin \theta \sin (\varphi -\Omega t)\;, 
\]
and at $r>R$ it becomes 
\[
\xi _{t}^{(+)}-\frac{2R^{3}}{5r^{2}}\mu \Omega \sin \theta \sin (\varphi
-\Omega t)\;, 
\]
\[
\xi _{r}^{(+)}=-2\frac{R}{r}\mu \sin \theta \cos (\varphi -\Omega t)\;, 
\]
\[
\xi _{\theta }^{(+)}=-2R\mu \cos \theta \cos (\varphi -\Omega t)\;, 
\]
\[
\xi _{\varphi }^{(+)}=2R\mu \sin \theta \sin (\varphi -\Omega t)\;. 
\]

\section{The $l=1$ metric perturbations}

Once the actual gauge vector $\xi _{\mu }^{(\pm )}$ is known, the MP are
constructed via Eq. (\ref{hpm}). Here we give the resultant expressions for $%
h_{\mu \nu }^{(\pm )}$, first at the leading order $(R\Omega )^{0}$, and
then at the next order $(R\Omega )^{1}$ (next subsection). Notice that all
three time-space components $h_{tr}^{(\pm )},h_{t\theta }^{(\pm
)},h_{t\varphi }^{(\pm )}$ vanish at the leading order. The magnitude of
these three components is $\propto \Omega $. Notice also that $h_{r\theta }$
and $h_{r\varphi }$ are smooth across $r=R$, and $h_{\theta \varphi }$
vanishes at both sides.

\subsection{Leading order ($\Omega ^{0}$) metric perturbations}

Here are the values of $h_{\mu \nu }^{(\pm )}$ at the leading order $
(M/R)^{0}=(R\Omega )^{0}$.

\textbf{Internal metric perturbations (}$2M<<r<R$\textbf{)} 
\[
h_{tt}^{(-)}=2\mu \left( \frac{r}{R^{2}}-\frac{R}{r^{2}}\right) \sin \theta
\cos (\varphi -\Omega t), 
\]
\[
h_{rr}^{(-)}=2\mu \left( \frac{r}{R^{2}}+\frac{R}{r^{2}}\right) \sin \theta
\cos (\varphi -\Omega t), 
\]
\[
h_{\theta \theta }^{(-)}=2\mu r^{2}\left( \frac{r}{R^{2}}-\frac{R}{r^{2}}
\right) \sin \theta \cos (\varphi -\Omega t), 
\]
\[
h_{\varphi \varphi }^{(-)}=2\mu r^{2}\left( \frac{r}{R^{2}}-\frac{R}{r^{2}}
\right) (\sin \theta )^{3}\cos (\varphi -\Omega t), 
\]
\[
h_{r\theta }^{(-)}=2\mu \frac{R}{r}\cos \theta \cos (\varphi -\Omega t), 
\]
\[
h_{r\varphi }^{(-)}=-2\mu \frac{R}{r}\sin \theta \sin (\varphi -\Omega t), 
\]
\[
h_{\theta \varphi }^{(-)}=0, 
\]
\[
h_{tr}^{(-)}=h_{t\theta }^{(-)}=h_{t\varphi }^{(-)}=0. 
\]
\textbf{External metric perturbations (}$R<r<<1/\Omega $\textbf{)}

\[
h_{rr}^{(+)}=4\mu \frac{R}{r^{2}}\sin \theta \cos (\varphi -\Omega t), 
\]
\[
h_{r\theta }^{(+)}=2\mu \frac{R}{r}\cos \theta \cos (\varphi -\Omega t), 
\]
\[
h_{r\varphi }^{(+)}=-2\mu \frac{R}{r}\sin \theta \sin (\varphi -\Omega t), 
\]
\[
h_{tt}^{(+)}=h_{\theta \theta }^{(+)}=h_{\varphi \varphi }^{(+)}=h_{\theta
\varphi }^{(+)}=0, 
\]
\[
h_{tr}^{(+)}=h_{t\theta }^{(+)}=h_{t\varphi }^{(+)}=0. 
\]

\subsection{Next-order ($\Omega ^{1}$) metric perturbations}

Here we give the expressions of the three time-space components $%
h_{tr}^{(\pm )},h_{t\theta }^{(\pm )},h_{t\varphi }^{(\pm )}$, which are all
proportional to $\Omega $. All other MP components have vanishing
contribution at this order, at both $r<R$ and $r>R$.

\textbf{Internal metric perturbations (}$2M<<r<R$\textbf{)}

\[
h_{tr}^{(-)}=2\mu \Omega \left( \frac{2r^{3}-10R^{2}r+5R^{3}}{5rR^{2}}
\right) \sin \theta \sin (\varphi -\Omega t), 
\]
\[
h_{t\theta }^{(-)}=-2\mu \Omega \left( \frac{r^{3}+10R^{2}r-5R^{3}}{5R^{2}}
\right) \cos \theta \sin (\varphi -\Omega t), 
\]
\[
h_{t\varphi }^{(-)}=-2\mu \Omega \left( \frac{r^{3}+10R^{2}r-5R^{3}}{5R^{2}}%
\right) \sin \theta \cos (\varphi -\Omega t). 
\]
\textbf{External metric perturbations (}$R<r<<1/\Omega $\textbf{)}

\[
h_{tr}^{(+)}=-2\mu \Omega R\left( \frac{5r^{2}-2R^{2}}{5r^{3}}\right) \sin
\theta \sin (\varphi -\Omega t), 
\]
\[
h_{t\theta }^{(+)}=-2\mu \Omega R\left( \frac{5r^{2}+R^{2}}{5r^{2}}\right)
\cos \theta \sin (\varphi -\Omega t), 
\]
\[
h_{t\varphi }^{(+)}=-2\mu \Omega R\left( \frac{5r^{2}+R^{2}}{5r^{2}}\right)
\sin \theta \cos (\varphi -\Omega t). 
\]

\subsection{Checking the above results}

The above expressions for $h_{\mu \nu }^{(\pm )}$ may be directly verified
by applying the following checks:

(a) The gauge condition $\bar{h}_{;\nu }^{\mu \nu }=0$ is satisfied at both $%
r>R$ and $r<R$; \footnote{%
The validity of this gauge condition at $r=R$ automatically follows from the
continuity of the MP there, because the gauge condition only includes
first-order derivatives.}

(b) The vacuum field equation $\Box h_{\mu \nu }=0$ is satisfied at both $%
r>R $ and $r<R$;

(c) The MP $h_{\mu \nu }^{(\pm )}$ is continuous at $r=R$;

(iv) The jump in $\bar{h}_{\mu \nu ,r}^{(\pm )}$ agrees with Eqs. (\ref{j1}-%
\ref{j3}) above.

Note that since we are only considering here the MP up to order $\Omega ^{1}$%
, when implementing the checks (a,b) we may pretend that spacetime is flat,
i.e. $M=0$.

We should emphasize that the above checks (a-d) by themselves do not fully
guarantee the validity of the derived MP, because they do not address the
issue of correct asymptotic behavior at the EH. (In our construction, this
issue is addressed by using the gauge vectors $\xi _{\mu }^{(\pm )}$ with
the correct asymptotic behavior at the EH and at infinity.)

\section{The overall metric perturbation}

In this section we explore the structure of the overall harmonic-gauge MP,
obtained by summing over all the tensor-harmonic modes. We shall consider
here the strict weak-field limit, i.e. the leading order $(M/R)^{0}=(R\Omega
)^{0}$ only. We then compare it to the trivial flat-space MP solution.

Consider first the flat-space value of the even $l=1$ harmonic-gauge MP. The
particle is now assumed to be static (corresponding to $\Omega =0$), located
at 
\[
r=R,\theta =\pi /2,\varphi =\varphi _{0}. 
\]
A straightforward calculation of the flat-space $l=1$ MP yields 
\[
h_{tt}^{(-)0}=h_{rr}^{(-)0}=2\mu \frac{r}{R^{2}}\sin \theta \cos (\varphi
-\varphi _{0}), 
\]
\[
h_{\theta \theta }^{(-)0}=2\mu \frac{r^{3}}{R^{2}}\sin \theta \cos (\varphi
-\varphi _{0}), 
\]
\[
h_{\varphi \varphi }^{(-)0}=2\mu \frac{r^{3}}{R^{2}}(\sin \theta )^{3}\cos
(\varphi -\varphi _{0}), 
\]
for $r<R$, and 
\[
h_{tt}^{(+)0}=h_{rr}^{(+)0}=2\mu \frac{R}{r^{2}}\sin \theta \cos (\varphi
-\varphi _{0}), 
\]
\[
h_{\theta \theta }^{(+)0}=2\mu R\sin \theta \cos (\varphi -\varphi _{0}), 
\]
\[
h_{\varphi \varphi }^{(+)0}=2\mu R(\sin \theta )^{3}\cos (\varphi -\varphi
_{0}), 
\]
for $r>R$. The label ''$0$'' was introduced here to denote the flat-space
quantities. All other components of $h_{\mu \nu }^{(\pm )0}$ vanish.

Let us denote by $\Delta h_{\mu \nu }$ the difference between the above
flat-space expressions for the even $l=1$ mode, and the corresponding
expressions of section VI for the Schwarzschild's weak-field limit. One
finds (after substituting $\Omega t\to \varphi _{0}$) 
\begin{equation}
\Delta h_{tt}=-2\mu \frac{R}{r^{2}}\sin \theta \cos (\varphi -\varphi _{0}),
\label{deltatt}
\end{equation}
\begin{equation}
\Delta h_{rr}=2\mu \frac{R}{r^{2}}\sin \theta \cos (\varphi -\varphi _{0}),
\label{deltarr}
\end{equation}
\begin{equation}
\Delta h_{\theta \theta }=-2\mu R\sin \theta \cos (\varphi -\varphi _{0}),
\label{delta3}
\end{equation}
\begin{equation}
\Delta h_{\varphi \varphi }=-2\mu R(\sin \theta )^{3}\cos (\varphi -\varphi
_{0}),  \label{delta4}
\end{equation}
\begin{equation}
\Delta h_{r\theta }=2\mu \frac{R}{r}\cos \theta \cos (\varphi -\varphi _{0}),
\label{delta5}
\end{equation}
\begin{equation}
\Delta h_{r\varphi }=-2\mu \frac{R}{r}\sin \theta \sin (\varphi -\varphi
_{0}),  \label{delta6}
\end{equation}
\begin{equation}
\Delta h_{\theta \varphi }=\Delta h_{tr}=\Delta h_{t\theta }=\Delta
h_{t\varphi }=0.  \label{delta7}
\end{equation}
Note that although $h_{\mu \nu }^{0}$ and Schwarzschild's $h_{\mu \nu }$ are
non-smooth at $r=R$, their difference $\Delta h_{\mu \nu }$ is smooth there.
(This must be the case, because $\Delta h_{\mu \nu }$ is a homogeneous
solution.)

Consider now the sum over all modes. We shall \textit{assume} here that
apart from the even $l=1$ mode considered here, all modes of the
harmonic-gauge MP behave normally at the limit $M\to 0$. \footnote{%
It appears likely (though it still needs be proved) that all radiative modes 
$l\geq 2$ will behave normally at the weak-field limit. For a circular orbit
it can be shown that the $l=0$ mode is normal. It is likely that for a
circular orbit the odd $l=1$ mode is normal too.} (This assumption helps in
providing a simple picture of the overall harmonic-gauge MP, though our main
conclusions do not depend on it, as we discuss in the next section.) In the
flat-space case, upon summation over all modes one recovers the standard
flat-space expression for the MP. We denote this overall flat-space
harmonic-gauge MP by $h_{\mu \nu }^{0(tot)}$ (here ''$0$'' stands for flat
space, and ''tot'' for the total MP, i.e. the sum over all modes). In
Cartesian coordinates $x,y,z,t,$ for the flat background, $h_{\mu \nu
}^{0(tot)}$ takes the simple form 
\begin{equation}
h_{tt}^{0(tot)}=h_{xx}^{0(tot)}=h_{yy}^{0(tot)}=h_{zz}^{0(tot)}=\frac{2\mu }{%
d},  \label{hflat}
\end{equation}
with all mixed Cartesian components vanishing, where $d$ is the distance
between the particle and the evaluation point.

Consider now the summation over modes in the Schwarzschild case (in the
weak-field limit). We denote this sum by $h_{\mu \nu }^{(tot)}$. Since the
even $l=1$ mode is assumed to be the only anomalous one, $h_{\mu \nu
}^{(tot)}$ will deviate from $h_{\mu \nu }^{0(tot)}$ just by the amount the
Schwarzschild's even $l=1$ mode deviates from its flat-space counterpart,
namely, 
\begin{equation}
h_{\mu \nu }^{(tot)}=h_{\mu \nu }^{0(tot)}+\Delta h_{\mu \nu }.
\label{twoparts}
\end{equation}

We conclude that at the weak-field limit the overall harmonic-gauge MP is
the sum of two terms:

(i) The standard flat-space expression $h_{\mu \nu }^{0(tot)}$, given in Eq.
(\ref{hflat}). This is a spherically-symmetric component, centered around
the source particle, which decreases like $1/d$;

(ii) The anomalous component $\Delta h_{\mu \nu }$, given in Eqs. (\ref
{deltatt}-\ref{delta7}). This is an $l=1$ term centered around the BH and
oriented in the particle's direction. Note that $\Delta h_{\mu \nu }$ is 
\textit{independent of the BH mass}.

Equation (\ref{twoparts}) is based on the above assumption that the even $%
l=1 $ mode is the only anomalous one. In case this is not the case, any
additional anomalous mode will give its contribution to the anomalous
component $\Delta h_{\mu \nu }$. This is further discussed in the next
section.

\section{Discussion}

The term $\Delta h_{\mu \nu }$ in Eq. (\ref{twoparts}) demonstrates the
anomalous behavior of the harmonic-gauge MP in the weak-field limit. This
term is caused by the presence of the BH, but its magnitude is independent
of the BH's mass $M$. Basically this term scales like 
\begin{equation}
\Delta h_{\mu \nu }\sim \frac{\mu R}{r^{2}}  \label{dhgeneral}
\end{equation}
[see e.g. the $tt$ or $rr$ components, Eqs. (\ref{deltatt},\ref{deltarr})].
This demonstrates an interesting pathology: For fixed $M$ and $\mu $, and
fixed evaluation point in the neighborhood of the BH (though still in the
weak-field region $r>>M$), the MP \textit{increases} linearly with the
distance $R$ to the source (whereas naturally one would expect that the
perturbation should decrease when its source is taken far away).

Furthermore, for fixed $M$ $,\mu $, and $R$, when one gets closer to the BH
the anomalous term grows steadily as $r^{-2}$. This growth continues as long
as $r>>M$. When approaching closer to the strong-field region, $\Delta
h_{\mu \nu }$ will approach a value of order 
\begin{equation}
\Delta h_{\mu \nu }\sim \mu R/M^{2}.  \label{maxdh}
\end{equation}

Consequently, even if the particle has a very small (but finite) mass $\mu
<<M$, it will produce a large harmonic-gauge MP ($\Delta h_{\mu \nu }>>1$)
in the near weak-field neighborhood of the BH, if the particle is situated
at a distance $R>>R_{0}$, where $R_{0}\equiv M^{2}/\mu $. This may be
achieved by either increasing $R$, or by decreasing $M$.

In fact, if a finite-mass particle is situated at $R>>R_{0}$ it is no longer
justified to consider the harmonic-gauge $h_{\mu \nu }$ as a ''linear
perturbation'', even if $\mu <<M$, because $\Delta h_{\mu \nu }>>1$. The
entire perturbation approach thus breaks down in this case (provided that
one insists on using the harmonic gauge!). Note, however, that the harmonic
gauge can be used in a linear perturbation analysis for arbitrarily large $R$%
, if the mass $\mu $ is taken to be infinitesimal (which corresponds to $%
R_{0}\to \infty $).

Detweiler and Poisson \cite{dp} recently found that the gravitational self
force (in the harmonic gauge \cite{bo}) acting on a particle in a circular
orbit around a Schwarzschild BH has a ''Newtonian'' term, i.e. a term that
does not vanish at the limit $M\to 0$ (though its very existence
does depend on the presence of the BH). Since the term $h_{\mu \nu
}^{0(tot)} $ in the overall harmonic-gauge MP is strictly isotropic around
the particle (it is spherically symmetric), the ''Newtonian self force''
solely originates from the anomalous component $\Delta h_{\mu \nu }$. It is
thus given by the contribution of $\Delta h_{\mu \nu }$ to the effective
''gravitational force'' term 
\[
-\mu \delta \Gamma _{\mu \nu }^{\lambda }u^{\mu }u^{\nu }\cong -\mu \delta
\Gamma _{tt}^{\lambda }\,, 
\]
where $\delta \Gamma $ denotes the perturbation in the connection, 
\[
\delta \Gamma _{\mu \nu }^{\lambda }=\frac{1}{2}g^{\lambda \eta }(h_{\eta
\mu ;\nu }+h_{\eta \nu ;\mu }-h_{\mu \nu ;\eta }), 
\]
and we have used here the Newtonian limit $u^{\mu }\cong \delta _{t}^{\mu }$%
. This only has $r$ component in our case (all other components vanish as $
\Omega \to 0$). Substituting Eq. (\ref{deltatt}) in the standard
expression for $\delta \Gamma $ one finds the self force 
\begin{equation}
f_{self}^{r}=\,(\mu /2)\Delta h_{tt,r}=2\mu ^{2}/r^{2}.  \label{sf}
\end{equation}
This agrees with the analysis by Detweiler and Poisson \cite{dp}.

The estimates in this section and the previous one were based on the
assumption that the only anomalous mode is the even $l=1$ mode. Even if this
assumption is not satisfied, the pathological behavior described above still
holds. Each additional anomalous mode will add a corresponding term to $%
\Delta h_{\mu \nu }$. Since each mode has a different angular dependence, it
is not possible to cure the problematic terms (\ref{dhgeneral}) or (\ref
{maxdh}) by adding other modes. For a generic angular direction, the small-$%
r $ (but yet with $r>>M$) asymptotic behavior will either remain that of (%
\ref{dhgeneral})---or perhaps become even worse [if additional anomalous
modes exist which have a radial dependence more singular than $r^{-2}$ (this
appears very unlikely, however)]. Therefore, the pathological behavior of
the harmonic gauge at the weak-field limit---e.g. the break-down of the
linear analysis for $R>R_{0}$---is insensitive to the assumption made here.

Next we comment on the domain of applicability of the expressions for the
harmonic-gauge MP derived in sections VI and VII. Consider a circular orbit
with some particular radius $R>>M$ (we assume here $R<<R_{0}$, so the linear
analysis is applicable). For too small $r$ values, various terms
proportional to $M/r$ will become important. For too large $r$ values,
various retardation effects may become important. Therefore the domain of
validity of the expressions derived in sections VI and VII is $%
M<<r<<1/\Omega $. Note that in the limit $M\to 0$ considered here, $%
\Omega $ vanishes too, hence this range covers the entire $r>0$ domain.

Throughout this paper we restricted attention to circular orbits. We expect,
however, that the main qualitative results---the anomaly of the MP, the
presence of ''Newtonian self force'', and also the pathological behavior as $%
R>R_{0}$---will apply to other types of weak-field orbits as well. The
explicit form of $\Delta h_{\mu \nu }$ may depend on the orbit under
consideration. In particular, for a non-circular orbit we do expect the
anomalous component $\Delta h_{\mu \nu }$ to include a contribution from $%
l=0 $ as well (however, the $l=0$ anomalous term is expected to be less
singular than that of the even $l=1$ mode, $\propto r^{-2}$). The form of $%
\Delta h_{\mu \nu }$ for non-circular orbits still need be explored.

\section*{Acknowledgment}

I am grateful to Misao Sasaki, Takahiro Tanaka, and all organizers of the
Sixth Capra Meeting and the successive discussion meeting, for the warm
hospitality. I also thank the participants of these meetings, and especially
Eric Poisson and Steve Detweiler, for many helpful discussions. This
research was supported by The Israel Science Foundation (grant no.
74/02-11.1). It was also supported in part by the Monbukagaku-sho
Grant-in-Aid for Scientific Research Nos. 14047212 and 14047214, and by the
Yukawa Institute for Theoretical Physics, Kyoto University.

\appendix

\label{A}
%

\section{} 

We explicitly give here five of the six even, $l=1$, exact static solutions
for Eq. (\ref{box}) in Schwarzschild spacetime. (The radial function of the
last solution, $\xi _{\mu }^{(C+)}$, will not be presented here as it is too
long.)

We first give the set of six modes with azimuthal number $m=0$ (i.e. the
axisymmetric modes). The first pair of modes, denoted $\xi _{\mu }^{(A\pm )}$%
, is given by 
\begin{equation}
\xi _{\mu }^{(A\pm )}=\phi _{;\mu }^{(\pm )}  \label{exactA}
\end{equation}
with 
\[
\phi ^{(\pm )}=A^{(\pm )}(r)\cos \theta , 
\]
where 
\begin{equation}
A^{(-)}(r)=r-M
\end{equation}
and 
\[
A^{(+)}(r)=(3/2)\frac{-2M+(M-r)\ln (1-2M/r)}{M^{3}}. 
\]
\footnote{%
The factor $3/2$, as well as the factor $3/4$ in the expression for $B^{(+)}$
below, where introduced in order for these expressions to coincide with the
corresponding expressions in section II in the weak-field limit.} Next, $\xi
_{\mu }^{(B\pm )}$ are given by 
\begin{equation}
\xi _{\mu }^{(B\pm )}=\delta _{\mu }^{t}B^{(\pm )}(r)\cos \theta ,
\end{equation}
where 
\begin{equation}
B^{(-)}(r)=r-2M\;
\end{equation}
and

\[
\;B^{(+)}(r)=(3/4)\frac{2M(r-M)+r(r-2M)\ln (1-2M/r)}{M^{3}r}. 
\]
Finally, the pair $\xi _{\mu }^{(C\pm )}$ takes the form 
\[
\xi _{r}^{(C\pm )}=C_{r}^{(\pm )}(r)\cos \theta , 
\]
\[
\xi _{\theta }^{(C\pm )}=C_{\theta }^{(\pm )}(r)\sin \theta , 
\]
\[
\xi _{\varphi }^{(C\pm )}=\xi _{t}^{(C\pm )}=0. 
\]
The radial functions $C_{r}^{(-)},C_{\theta }^{(-)}$ are 
\[
C_{r}^{(-)}(r)=\frac{-4M^{3}+3Mr^{2}+6r^{3}+4M^{2}r\ln (r/M)}{6r} 
\]
and 
\[
C_{\theta }^{(-)}(r)=\frac{12M^{3}+8M^{2}r-3Mr^{2}+6r^{3}+8M^{2}(M-r)\ln
(r/M)}{12}. 
\]
The remaining two radial functions $C_{r}^{(+)},C_{\theta }^{(+)}$ are much
more complicate, and we shall not present them here. Both functions (which
we have encoded in a MATHEMATICA file) are non-elementary, as they include
the function Polylog$[2,r/2M]$.

The other two sets of six modes corresponding to $m=\pm 1$ are obtained by
straightforward rotations. One of these sets, the one which is mostly
relevant to our analysis, is the following. The first pair $\xi _{\mu
}^{(A\pm )}$ is given by Eq. (\ref{exactA}), this time with 
\[
\phi ^{(\pm )}=A^{(\pm )}(r)\sin \theta \cos \varphi 
\]
($A^{(\pm )}$ as well as $B^{(\pm )}$ and $C_{\mu }^{(\pm )}$ below are the
same as specified above). The second pair is 
\begin{equation}
\xi _{\mu }^{(B\pm )}=\delta _{\mu }^{t}B^{(\pm )}(r)\sin \theta \sin
\varphi .  \label{exactt}
\end{equation}
The last pair $\xi _{\mu }^{(C\pm )}$ now takes the form 
\[
\xi _{r}^{(C\pm )}=C_{r}^{(\pm )}(r)\sin \theta \cos \varphi \;, 
\]
\[
\xi _{\theta }^{(C\pm )}=-C_{\theta }^{(\pm )}(r)\cos \theta \cos \varphi
\;, 
\]
\[
\xi _{\varphi }^{(C\pm )}=C_{\theta }^{(\pm )}(r)\sin \theta \sin \varphi
\;, 
\]
\[
\xi _{t}^{(C\pm )}=0. 
\]

The last set of six $m\neq 0$ modes is obtained by transforming $\varphi
\to (\varphi +\pi /2)$.

\end{document}